\documentclass[11pt]{article}
\usepackage{DGfest,epsfig}
\usepackage{times,bm,amsfonts,amssymb}

\def\Journal#1#2#3#4{{#1} {\bf #2}, #3 (#4)}

\def\JHEP{\em JHEP}

\def\NPB{{\em Nucl. Phys.} B}
\def\PLB{{\em Phys. Lett.}  B}
\def\PLA{{\em Phys. Lett.}  A}
\def\PRL{\em Phys. Rev. Lett.}
\def\PRD{{\em Phys. Rev.} D}
\def\RMP{Rev. Mod. Phys.}

\def\be{\begin{equation}}
\def\ee{\end{equation}}
\def\bea{\begin{eqnarray}}
\def\eea{\end{eqnarray}}

\begin{document}
\baselineskip 11.5pt
\title{DEPARTURES FROM SPECIAL RELATIVITY BEYOND
EFFECTIVE FIELD THEORIES}

\author{J.M. CARMONA}
\address{Departamento de F\'{\i}sica Te\'orica, Universidad de Zaragoza,
50009 Zaragoza, Spain \\
Instituto de Biocomputaci\'on y  F\'{\i}sica de
Sistemas Complejos (BIFI), 50009 Zaragoza, Spain}

\author{J.L. CORT\'ES}
\address{Departamento de F\'{\i}sica Te\'orica, Universidad de Zaragoza,
50009 Zaragoza, Spain}

%

\maketitle\abstracts{The possibility to have a deviation from
relativistic quantum field theory requiring to go beyond effective
field theories is discussed. A few recent attempts to go in this
direction both at the theoretical and phenomenological levels are
briefly reviewed.}

\section{Introduction}

Lorentz invariance is a fundamental ingredient of our present
description of Nature, which is given in terms of relativistic
quantum field theories (RQFTs). However, the idea that Lorentz
invariance might be an approximate, low-energy, symmetry, has begun
to emerge in the last few years essentially from quantum gravity
developments~\cite{qugr}, but also from results in the fields of
string theory~\cite{str}, noncommutative geometry~\cite{ncprl}, or
other ideas such as varying couplings~\cite{vcoupl}.

There are other motivations to think on limitations of the RQFT
framework: the strong difficulties in obtaining a RQFT containing
gravitation, the large mismatch between the expected and measured
value of the energy density of vacuum (cosmological constant
problem), or the consideration that in a quantum theory of gravity
the maximum entropy of any system should be proportional to the area
and not to the volume~\cite{holo}, contrary to what happens in
RQFT~\cite{holoQFT}.

There exist also some phenomenological results that could find a
simple explanation in terms of a Lorentz invariance violation
(LIV). The observation of cosmic rays with energies above $5\times
10^{19}$\,eV seems to violate the GZK cutoff~\cite{gzk}, whose
existence is implied by relativistic kinematics. On the other
hand, CPT symmetry might also be violated if Lorentz symmetry were
not exact, which could provide the key to explain the
matter-antimatter asymmetry of our Universe~\cite{CPTmatter}.

The framework of effective field theories provides a conservative
approach to incorporate departures from special relativity. Here
Lorentz symmetry is spontaneously broken, which produces the
apparition of Lorentz non-invariant terms in an effective
Lagrangian. This approach has been extensively studied in the last
years~\cite{kostel}. It assumes that one can incorporate the
corrections order by order in the effective theory, so that, for a
certain level of precision, the effective Lagrangian has always a
finite number of terms. Tests of special relativity have put
strong bounds on different terms of the effective
Lagrangian~\cite{phystoday}.

In this paper we want to point out that the effective field theory
framework may be too tight to include corrections coming from a
LIV, giving explicit examples of theoretical schemes beyond that
framework.

\section{Beyond effective field theories}

\subsection{A fundamental theory}

With the exception of the attempts to consider a renormalizable
theory of gravity at the nonperturbative level~\cite{safety}, all
other approaches to a fundamental theory of quantum gravity (QG) go
beyond QFT. This includes string theory~\cite{st} and their related
ideas like a four dimensional theory immersed in a non trivial way in
higher dimensions~\cite{branes}. It is generally assumed that the
success of RQFT as a theory of the fundamental interactions of
particles (with the exception of gravity) is a consequence of a
decoupling between the degrees of freedom which are incorporated in
QFT and all the remaining degrees of freedom of the fundamental theory
whose virtual effects can be incorporated at the level of QFT. But it
might be that this is not the case and some traces remain at low
energies which can not be incorporated in a QFT. Our present
understanding of the properties of a theory of QG does
not allow to exclude this possibility.

The most direct approach would be to look for possible candidates
for the fundamental theory and in each case see whether the low
energy limit can be described by a QFT. But we are far from the
identification of the fundamental theory and a systematic derivation
of its low energy limit. It seems then reasonable to explore
extensions of QFT as candidates for such low energy limit.

\subsection{Non-commutative quantum field theory}

A first indication of the possible limitation of a discussion of
Lorentz violating effects within the framework of effective field
theories comes from the study of noncommutative QFT.

A non-commutativity of space-time has been considered as a possible
signature of QG. Then it is natural to consider the
formulation of a QFT in a noncommutative space-time~\cite{Nekrasov}
as a possible framework for departures from conventional QFT induced
by gravitational effects.

A generic feature is the appearance of Lorentz violation effects.
One can make explicit computations in different models to identify
the quantum effects induced by the non-commutativity. A surprise of
this analysis is the appearance of an infrared/ultraviolet (IR/UV)
mixing~\cite{IRUV},
i.e., a dependence of the low energy effective action on the
ultraviolet scale $M$ introduced to regulate the UV divergences in
loop diagrams. As a consequence of this mixing there is an ambiguity
in the estimates of signals of the non-commutativity at low
energies. In fact the commutative limit and the low energy limit
($M\to \infty$) do not commute. Depending on how one approaches to the
conventional QFT limit one has different results. It may be that the
IR/UV mixing has no physical consequences and the effects of a
noncommutative space-time can be incorporated at the level of
effective field theories~\cite{ncprl} or alternatively one can have
a remaining non-locality due to extended degrees of freedom at low
energies which can not be incorporated in the effective field theory
framework~\cite{BanksAmelino}. Which of these cases (if any) is
realized in QG will depend on the details of the
underlying theory.

\subsection{Quantum theory of non-commutative fields}

Another way to go beyond conventional QFT is based on the
introduction of a non-commutativity in the space of field
configurations~\cite{qncft}. The canonical commutation relations
between the field and the conjugated momentum at each space point is
supplemented by a non trivial commutator between different fields
and/or between different conjugated momenta. When these generalized
commutation relations are combined with the conventional field
theory Hamiltonian one has a generalization of QFT with a violation
of Lorentz invariance.

The simplest model with non-commutative fields~\cite{qncft} is the
free theory of two scalar fields. There are two different type of
excitations (the particle-antiparticle symmetry is broken by the
non-commutativity of the fields). Two energy scales parametrize the
generalization of the commutation relations. When the ratio of these
two energy scales is very small there is a domain of energies
between the two scales where one approaches the conventional theory
of relativistic particles and antiparticles.

In order to illustrate the dynamical consequences of the
non-commutativity of fields one can consider the Hamiltonian
formulation of non-commutative scalar QED. One has a Hamiltonian
\begin{equation}
H \,=\, H_g + H_m
\end{equation}
where
\begin{equation}
 H_g \,=\, \frac{1}{2} {\bf E}^2 + \frac{1}{2}
\left({\bf \nabla} \wedge {\bf A}\right)^2
\end{equation}
is the Hamiltonian of the electromagnetic field with commutators
\begin{equation}
\left[A_j({\bf x}), E_k({\bf y})\right] \,=\, i \delta_{jk}
\delta\left({\bf x}-{\bf y}\right)
\end{equation}
and
\begin{equation}
H_m \,=\, \Pi^{\dagger} \Pi + \left({\bf \nabla}\Phi^{\dagger} + i
{\bf A} \Phi^{\dagger}\right) \left({\bf \nabla}\Phi - i {\bf A}
\Phi\right) + m^2 \Phi^{\dagger} \Phi
\label{Hm}
\end{equation}
is the Hamiltonian of the matter system corresponding to the theory
of a complex non-commutative scalar field with commutators
\begin{equation}
\left[\Phi({\bf x}), \Pi^{\dagger}({\bf y})\right] \,=\, -
\left[\Pi({\bf x}), \Phi^{\dagger}({\bf y})\right] \,=\, i
\delta\left({\bf x}-{\bf y}\right)
\end{equation}
\begin{equation}
\left[\Phi({\bf x}), \Phi^{\dagger}({\bf y})\right] \,=\,
\frac{1}{\Lambda}\, \delta\left({\bf x}-{\bf y}\right)
\end{equation}
\begin{equation}
\left[\Pi({\bf x}), \Pi^{\dagger}({\bf y})\right] \,=\, \lambda\,
\delta\left({\bf x}-{\bf y}\right)
\end{equation}
Physical states satisfy the constraint
\begin{equation}
\left[{\bf \nabla} {\bf E} ({\bf x}) - \rho ({\bf x})\right]
\,|\Psi\rangle_{phys} \,=\, 0
\end{equation}
with
\begin{equation}
Q \,=\, \int d^3{\bf x}\, \rho ({\bf x})
\label{Qrho}
\end{equation}
the generator of $U(1)$ transformations of the scalar field.

It is straightforward to go from the Hamiltonian to the
Lagrangian formulation following step by step the case
of conventional scalar QED.
The final result is a Lagrangian
\begin{equation}
{\cal L} \,=\, {\cal L}_g + {\cal L}_m
\end{equation}
with
\begin{equation}
{\cal L}_g \,=\, \frac{1}{2} \left(\partial_t {\bf A} - {\bf \nabla}
A_0\right) \left(\partial_t {\bf A} - {\bf \nabla} A_0\right) -
\frac{1}{2} \left({\bf \nabla}\wedge {\bf A}\right) \left({\bf
  \nabla}\wedge {\bf A}\right)
\end{equation}
which is the conventional Lagrangian of the electromagnetic field
(non-commutativity is introduced in the matter sector). The modified
Lagrangian of the matter system is given by
\begin{eqnarray}
{\cal L}_m &=& - \left(\frac{1}{1-\lambda/\Lambda}\right)^2 \,
\Phi^{\dagger} \frac{\left(\partial_t - i A_0\right)^2}{1-i
\left(\partial_t - i
  A_0\right)/\left(\Lambda-\lambda\right)}  \Phi
\nonumber \\ && + \, \Phi^{\dagger} \left(\nabla -i {\bf A}\right)^2
\Phi - m^2 \Phi^{\dagger} \Phi
\end{eqnarray}
One can see that the non-commutativity translates into a
non-locality in the Lagrangian which makes manifest how the theory
of non-commutative fields goes beyond the effective field theory
approach.

\subsection{New infrared scales}

Effective field theories have a limited range of applicability. They
are supposed to give sensible descriptions at energies $E$ much
lower than a high-energy scale $M$, the ultraviolet cutoff of the
theory, whose effect can be incorporated by nonrenormalizable terms
which produce corrections of order $(E/M)$. The scale $M$
corresponds usually to the mass of a very massive particle, so that
the introduction of this energy scale does not pose any problems
with respect to relativistic invariance.

On the other hand, the introduction of corrections to a quantum
field theory parametrized by a low-energy scale has not been so well
explored in the literature. In fact, depending on how this energy
scale is introduced, these corrections could violate relativistic
invariance. There are, however, several phenomenological and
theoretical reasons that lead to think on the necessity to
incorporate a new IR scale to our theories. These include:

\begin{enumerate}

\item The seeming existence of a cosmological constant or a vacuum
energy density whose experimental value, $\rho_V \sim
(10^{-3}\,\mbox{eV})^4$ is $124$ orders of magnitude lower than its
expected value from ordinary quantum field theory~\cite{dsrir}.

\item The fact that the entropy in a field theory scales with the
volume, while in a quantum theory of gravity the maximum entropy
should be proportional to the area leads to think that conventional
quantum field theories overcount degrees of freedom~\cite{holo},
which suggests the breakdown of any effective theory with an
ultraviolet cutoff to describe systems which exceed a certain
critical size which depends on the ultraviolet
cutoff~\cite{holoQFT}. This critical size constitutes an IR
energy scale.

\item In the approach of large extra dimensions~\cite{extradim},
the observed hierarchy between the electroweak and Planck scales
is explained by postulating a fundamental scale $M\sim 10 - 100$
TeV of gravity along with Kaluza-Klein compactification with large
radius $R$, so that the Planck scale is then an effective
four-dimensional scale. The inverse of $R$ is an IR energy scale.

\end{enumerate}

A deviation from a RQFT at low energies due to an IR
scale $\lambda$ may well be expected
to violate relativistic invariance. The reason is that it has been
shown~\cite{weinberg} that {\it any} theory incorporating quantum mechanics
and special relativity, with an additional ``cluster''
condition~\cite{weinberg}, must reduce to a RQFT at low energies.

A simple way of incorporating effects beyond RQFT that violate
relativistic invariance is through a modified dispersion relation.
For example, the dispersion relation
\begin{equation}
E^2 = \bm{p}^2 + m^2 + \lambda |\bm{p}| ,
\label{lorentz1}
\end{equation}
has a term linear in $|\bm{p}|$ which dominates over the standard
kinetic term $(\bm{p}^2)$ when
$|\bm{p}|\lesssim 2\lambda$ and so it changes drastically
the nonrelativistic kinematics.
For instance, such a dispersion relation for the electron
would slightly modify the energy levels of the hydrogen atom.
Given the extraordinary agreement between theory
and the experimental measurement of the Lamb shift (one part in
$10^5$~\cite{weinberg,kinoshita}), one has $\lambda<10^{-6}-10^{-7}$ eV,
which is a very stringent bound on the IR scale.

The bound is much less restrictive and more interesting if one considers
that Eq.~(\ref{lorentz1}) applies to the neutrino only because of a
special sensitivity of this particle to the IR scale. This idea
could find theoretical support in the framework of large extra dimensions,
considering that the neutrino is the only particle, together with the
graviton, which propagates in the extra dimensions, and that the dependence
on the scale $\lambda$ is suppressed for the remaining particles,
which would then explain why no signal of these LIVs has been observed.
The neutrino has two characteristic properties: it has a very small
mass, and it interacts only weakly. As a result of this combination, we have
not any experimental result on its nonrelativistic physics. Therefore, the
presence of Lorentz invariance violations affecting the nonrelativistic limit
cannot be excluded a priori in the neutrino case.

In fact Eq.~(\ref{lorentz1}) for the neutrino has been used~\cite{tritio} to
explain the tritium beta-decay anomaly, which consists in an excess of electron
events at the end of the
spectrum, at about 20 eV below the end point~\cite{lobashev}.
Matching with experimental results requires a value of $\lambda$ in
Eq.~(\ref{lorentz1}) of the order of the eV.
It can be seen~\cite{tritio} that this does not contradict other experimental
results involving neutrinos, such as their contribution to the
energy density of the Universe, neutrino oscillations (if the IR
scale is family-independent) or neutrinos from supernovas, although
this could be a good place to look for footprints of this LIV.

Eq.~(\ref{lorentz1}) for the neutrino might however have
important effects in cosmic rays through threshold effects which become relevant when
$\lambda |{\bm p}_{\rm th}|\sim m^2$. Here $m^2$ is an ``effective''
mass squared which controls the kinematic condition of allowance or
prohibition of an specific process. Indeed a consequence of
these threshold effects could be that neutrons and pions become
stable particles at energies close to the knee of the cosmic ray
spectrum~\cite{tritio}, which would drastically alter the composition
of cosmic rays. It is quite remarkable that cosmic ray phenomenology
could be sensitive to the presence of an IR scale.

A modified dispersion relation of the form of Eq.~(\ref{lorentz1})
cannot be simply introduced in the framework of an effective
field theory, because the term proportional to $|\bm{p}|$
cannot be Taylor-expanded. We therefore lack of a dynamical formalism
consistent with such a dispersion relation, and we can only explore
for the moment its kinematic implications.

There is another difficulty of introducing a LIV through
a modified dispersion relation, that is,
one should indicate the ``preferred'' frame in which this relation
is valid. There is however another possibility, which is to extend our
concept of relativistic invariance to a more general framework, in
which the new dispersion relation would be observer-invariant.
This case is considered in the following section.

\subsection{Double Special Relativity}

The possibility that Lorentz symmetry, considered as a low-energy symmetry
that would not be exactly preserved in a quantum theory of gravity,
might not be broken in the fundamental theory, but only deformed to
a different symmetry, was started to be explored quite recently~\cite{dsr1}.
This deformation usually involves~\cite{dsr2}
a new dispersion relation of the form
$m^2=f(E,p;M)$, with $f(E,p;M)\to E^2-p^2$ in the $M\to \infty$ limit,
where $M$ is a large-energy/small-length (possibly
related to the Planck mass) scale, which is introduced as an
observer-independent scale. The new dispersion relation then implies
new laws of boost/rotation transformation between inertial observers.
As Galilean Relativity was deformed to Special Relativity in order to
introduce a relativistic-invariant scale (the speed of light), the new
framework [therefore called \emph{Doubly Special Relativity} (DSR)]
deforms Special Relativity to introduce this second observer-independent
scale.

There is an special point that needs to be remarked in
the kinematical analysis of physical processes in the DSR framework.
The assumption of modified dispersion relations and unmodified laws
of energy-momentum conservation is inconsistent with the
doubly-special relativity principles, since it inevitably~\cite{dsr1}
gives rise to a preferred class of inertial observers. A doubly-special
relativity scenario with modified dispersion relations must
therefore necessarily have a modified law of energy-momentum conservation.

On the contrary, unmodified laws of energy-momentum conservation are
an ingredient of any low-energy effective field theory. This fact is
an indication that DSR theories are not included in the effective field
theory framework.

It is interesting to note that one can find a phenomenologically
consistent \emph{infrared} DSR~\cite{dsrir},
meaning the introduction of a new IR scale $\lambda$,
as discussed in the previous section, but in a
relativistic-invariant way. In particular, it is possible
to obtain appropriate deformed Lorentz transformations such
that the following dispersion relation
\begin{equation}
\tilde m^2=\left(1-\frac{\lambda}{E}\right)^2 (E^2-\bm{p}^2),
\end{equation}
remains invariant. In the nonrelativistic limit, this
dispersion relation is reduced to
\begin{equation}
E\simeq m+\frac{\bm{p}^2}{2m}-\frac{\tilde m}{m}
\frac{\bm{p}^4}{8m^3},
\end{equation}
where we have defined the physical mass parameter $m=\tilde
m+\lambda$. Tests of QED now give a bound for the IR scale
$\lambda\lesssim 10^{-2}\,$eV~\cite{dsrir}. The kinematical
dependence on the IR scale in the case of DSR is very different from
the one considered in the previous section. This is the reason why
the bounds on the IR scale from precision QED tests differ by
several orders of magnitude. Interestingly enough, a value of the
new scale of order $10^{-3}\,$eV could be detected in the near
future from its QED effects, and, at the same time, provide a
solution to the cosmological constant problem. Assuming a mechanism
of cancelation for the different contributions to the vacuum
expectation value of the energy momentum tensor, then on dimensional
grounds one would expect
\begin{equation}
\langle T_{\mu\nu}\rangle \sim \lambda^4 \left(c_1 \delta_{\mu}^0
\delta_{\nu}^0 + c_2 \sum_i \delta_{\mu}^i \delta_{\nu}^i\right)
\end{equation}
with $c_1$, $c_2$ dimensionless coefficients depending on the details
of the theory incorporating the new low energy scale. Then the present
acceleration of the expansion of the Universe could be a signal of a
new low energy scale compatible with a modified relativity principle.

\section{Conclusions}

We are entering in a period where an experimental search of QG effects
is possible. One of the most clean signals of these effects would be a
departure from special relativity, since
ultra-high energy cosmic rays and astrophysical observations
can provide amplification mechanisms (through kinematical thresholds,
modification of stability or instability conditions for high-energy
particles, or time-of-flight measurements of far enough energy sources)
by which these effects might be observable in a close future.
On the other hand, QED and/or neutrino physics could also detect the
presence of an IR scale producing a slight modification of the kinematics in the
nonrelativistic limit.

The details of the way one
approaches to a special relativistic theory will give us very
important hints on the underlying theory. Lacking the basic principles
of this theory one should keep an open mind and explore different
alternatives for extensions of RQFT considering also the possibility
to go beyond effective field theories. We have sketched a few of these
alternatives as candidates for a phenomenological perspective to the
analysis of QG effects.

\section*{Acknowledgments}
This work is supported by CYT FPA, grant 2003-02948 and by INFN-CYT
collaboration grant.

\end{document}